# The Application of Cloud Computing to the Creation of Image Mosaics and Management of Their Provenance

G. Bruce Berriman*[a], Ewa Deelman[b], Paul Groth[b], and Gideon Juve[b]
[a] Infrared Processing and Analysis Center, California Institute of Technology,
770 South Wilson Avenue, Pasadena, CA 91125, USA
[b] Information Sciences Institute, University of Southern California,
4676 Admiralty Way, Suite 1001, Marina del Rey, CA 90292, USA

## ABSTRACT

We have used the Montage image mosaic engine to investigate the cost and performance of processing images on the Amazon EC2 cloud, and to inform the requirements that higher-level products impose on provenance management technologies. We will present a detailed comparison of the performance of Montage on the cloud and on the Abe high performance cluster at the National Center for Supercomputing Applications (NCSA). Because Montage generates many intermediate products, we have used it to understand the science requirements that higher-level products impose on provenance management technologies. We describe experiments with provenance management technologies such as the "Provenance Aware Service Oriented Architecture" (PASOA).

**Keywords:** Image processing, image mosaics, cloud computing, computational performance, provenance, high performance clusters, astronomical images

## 1. INTRODUCTION

The late Jim Gray described how science is becoming "data-centric" and how information technology is transforming the way science is done [1]. Vast quantities of data are made available to scientists at an ever-accelerating rate, and sophisticated and innovative approaches to data discovery, data mining and analysis are being developed to extract the full scientific content of this data tsunami. This paradigm of "e-Science," as it has come to be known, is enabling the production of new data products, synthesized from existing data products and often produced by reprocessing or re-sampling data products.

We have investigated two aspects of data processing and management that have been raised by this broad theme of e-Science:

- What is the cheapest and most efficient way of processing data to create new products, given that data are distributed across the world and that processing them often requires access to high-end compute and storage resources? Should data be uploaded to the software, or should software be built close to the data?
- How can we capture the processing histories of these third-party products and capture their processing histories and provenances?

There are many powerful compute, storage and network technologies now available to scientists, such as high performance clusters, grids, parallel file systems; see [2] for a thorough account of the state of the art. Cloud computing is now receiving a lot of attention from scientists as a technology for running scientific applications. While the definition of cloud-computing tends to be in the eye of the beholder, in this context we consider the cloud as a cluster that offers virtualized services, service-oriented provisioning of resources and, most important of all, "pay-as-you-go" pricing. Cloud computing is now offered by a number of commercial providers, and Open Source clouds are also available. The benefits are obvious: end-users pay only for what they use, the strain on local system maintenance and power consumption is reduced, and resources are available on demand ("elasticity") and are not subject to scheduling overheads. Clouds are, however, built with commodity hardware and network devices, and generally do not offer the high-speed networks and parallel file systems that are installed in many high-end clusters and may be at a serious performance disadvantage for many applications.

* gbb@ipac.caltech.edu; phone 1 626 395-1817; fax 1 626 395-6666; astrocompute.wordpress.com

Early studies investigated the cost and performance of tightly coupled applications running on the cloud [3]. In this study, we focus on investigating the performance and cost of running workflow applications on the cloud, including a performance comparison with a high-performance cluster, and on investigating the cost of uploading and storing data in the cloud. Workflows in this context are loosely coupled "pipeline" applications where the output from one component is input to the next. We will concentrate on an astronomy workflow application, Montage, an I/O bound application, and compare its performance with two applications from other fields. A more thorough description of this work is given in [4].

As part of this study, we have conducted experiments in capturing provenance records from the Montage processing logs. This work remains experimental, but the results are sufficiently encouraging to merit discussion. Astronomers need to understand the technical content of data sets and evaluate published claims based on them. All data products and records from all the steps used to create science data sets ideally would be archived, but the volume of data would be prohibitively high. The high cadence surveys currently under development will exacerbate this problem; the Large Synoptic Survey Telescope alone is expected to deliver 60 PB of just raw data in its operational lifetime. There is therefore a need to create records of how data were derived - provenance - that contain sufficient information to enable their replication. A report issued by the National Academy of Sciences dedicated to the integrity of digital data [5] and an NSF workshop on software sustainability [6] both recognized the importance of curation of the provenance of data sets and of the software creating them.

Provenance records must meet strict specifications if they are to have value in supporting research. They must capture the algorithms, software versions, parameters, input data sets, hardware components and computing environments. The records should be standardized and captured in a permanent store that can be queried by end users. In this paper, we describe how the Montage image mosaic engine acts as a driver for the application in astronomy of provenance management; see [7] for an overview of the subject.

## 2. EVALUATING APPLICATIONS ON THE AMAZON EC2 CLOUD: SET-UP

### 2.1 Goals of this Study

Briefly, our goal is to determine which types of *scientific workflow applications* are cheaply and efficiently run on the Amazon EC2 cloud (hereafter, AmEC2). Workflows are loosely coupled parallel applications that consist of a set of computational tasks linked via data- and control-flow dependencies. Unlike tightly coupled applications, in which tasks communicate directly via the network, workflow tasks typically communicate using the file system. Given that AmEC2 uses only commodity hardware and given that applications make very different demands on resources, it is likely that cost and performance will vary dramatically with application. It was therefore important to study applications that make different demands on resources. Thus the goals of this study are

- Understand the performance of three workflow applications with different I/O, memory and CPU requirements on a commercial cloud
- Compare the performance of the cloud with that of a high performance cluster (HPC) equipped with high-performance networks and parallel file systems, and
- Analyze the various costs associated with running workflows on a commercial cloud.

### 2.2 The Workflow Applications

We have chosen three workflow applications because their usage of computational resources is very different: **Montage** (http://montage.ipac.caltech.edu) from astronomy, **Broadband** (http://scec.usc.edu/research/cme/), from seismology, **Epigenome** (http://epigenome.usc.edu/) from biochemistry. Because the intended audience of this paper is astronomers, we will describe the performance of Montage in greater detail than the other two. **Montage** (http://montage.ipac.caltech.edu) is a toolkit for aggregating astronomical images in Flexible Image Transport System (FITS) format into mosaics. Its scientific value derives from three features of its design:

- It uses algorithms that preserve the calibration and positional (astrometric) fidelity of the input images to deliver mosaics that meet user-specified parameters of projection, coordinates, and spatial scale. It supports all projections and coordinate systems in use in astronomy.
- It contains independent modules for analyzing the geometry of images on the sky, and for creating and managing mosaics.
- It is written in American National Standards Institute (ANSI)-compliant C, and is portable and scalable the same engine runs on desktop, cluster, supercomputer environments or clouds running common Unix-based operating systems.

There are four steps in the production of an image mosaic:
- Discover the geometry of the input images on the sky from the input FITS keywords and use it to calculate the geometry of the output mosaic on the sky.
- Re-project the input images to the spatial scale, coordinate system, World Coordinate System (WCS)-projection, and image rotation.
- Model the background radiation in the input images to achieve common flux scales and background level across the mosaic.
- Co-add the re-projected, background-corrected images into a mosaic.

Each production step has been coded as an independent engine run from an executive script. Figure 1 illustrates the second through fourth steps for the simple case of generating a mosaic from three input mosaics. In practice, as many input images as necessary can be processed in parallel, limited only by the available hardware.

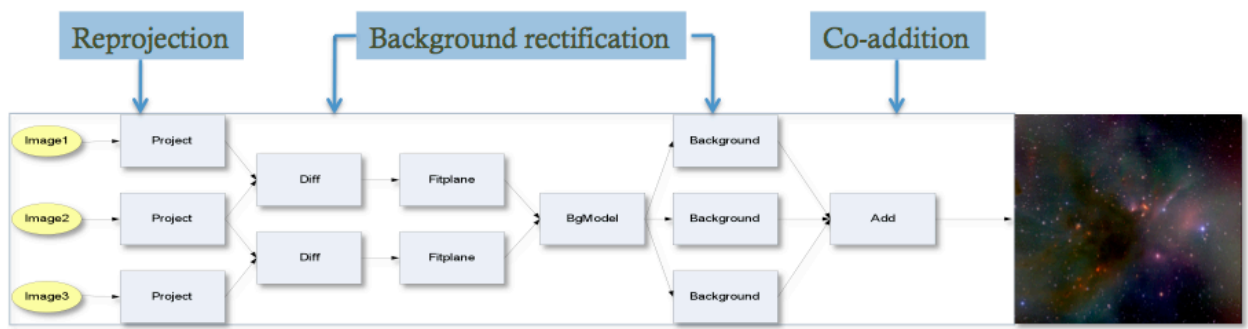

Figure 1: The processing steps used in computing an image mosaic with the Montage engine.

**Broadband** generates and compares seismograms from several high- and low-frequency earthquake simulation codes. Each workflow generates seismograms for several sources (earthquake scenarios) and sites (geographic locations). For each (source, site) combination the workflow runs several high- and low-frequency earthquake simulations and computes intensity measures of the resulting seismograms.

**Epigenome** maps short DNA segments collected using high-throughput gene sequencing machines to a previously constructed reference genome. The workflow splits several input segment files into small chunks, reformats and converts the chunks, maps the chunks to a reference genome, merges the mapped sequences into a single output map, and computes the sequence density for each location of interest in the reference genome.

### 2.3 Workflows Used In This Study

We configured a single workflow for each application throughout the study. Table 1 summarizes the resource usage of each of them, graded as 'high,' 'medium,' or 'low.'

**Montage** generated an 8-degree mosaic of M16 composed of images from the Two Micron All Sky Survey (2MASS; http://www.ipac.caltech.edu/2mass/). The resulting workflow contains 10,429 tasks, reads 4.2 GB of input data, and produces 7.9 GB of output data. Montage is considered I/O-bound because it spends more than 95% of its time waiting on I/O operations.

**Broadband** used 4 sources and 5 sites to generate a workflow containing 320 tasks that reads 6 GB of input data and writes 160 MB of output data. Broadband is considered to be memory-limited because more than 75% of its runtime is consumed by tasks requiring more than 1 GB of physical memory.

The **Epigenome** workflow maps human DNA sequences to a reference chromosome 21. The workflow contains 81 tasks, reads 1.8 GB of input data, and produces 300 MB of output data. Epigenome is CPU-bound because it spends 99% of its runtime in the CPU and only 1% on I/O and other activities.

Table 1: Application Resource Usage Comparison

| **Application** | **I/O** | **Memory** | **CPU** |
| --- | --- | --- | --- |
| Montage | High | Low | Low |
| Broadband | Medium | High | Medium |
| Epigenome | Low | Medium | High |

### 2.4 Experimental Set-Up and Execution Environment

We ran experiments on AmEC2 (http://aws.amazon.com/ec2/) and NCSA's Abe High Performance Cluster (HPC) (http://www.ncsa.illinois.edu/UserInfo/Resources/Hardware/Intel64Cluster/). AmEC2 is the most popular, feature-rich, and stable commercial cloud. Abe is typical of the existing HPC systems a scientist could choose for workflow applications, and it is equipped with high-speed networks and parallel file systems to provide high-performance I/O. These high-performance devices increase workflow performance by making inter-task communication more efficient. To have an unbiased comparison of the performance of workflows on AmEC2 and Abe, the experiments presented here run all experiments on single nodes, using the local disk on both EC2 and Abe, and the parallel file system on Abe. Intuitively, the parallel file system would be expected significantly improve the runtime of I/O-intensive applications like Montage, but will be less of an advantage for CPU-intensive applications like Epigenome.

**Compute Resources**

Table 2 lists the five AmEC2 compute resources chosen for this study, and we will refer to them by their AmEC2 designations *m1\** and *c1\**. These resources span the range of properties offered by AmEC2, and include 32-bit and 64-bit machines, with memories from 1.7 GB to 15 GB, and from 1 to 8 cores. The *m1* resources use Opteron processors, while the *c1* resources used Xeon processors, which have the superior floating-point performance of the two (4 floating-point operations/cycle vs. 2 floating-point operations/cycle).

The two Abe nodes use the same resource type – a 64-bit Xeon machine – but differ only in their I/O devices: *abe.local* uses a local partition for I/O, and the *abe.lustre* type uses a Lustre™ parallel-file partition (see http://en.wikipedia.org/wiki/Lustre_(file_system)). Both instances use a 10-Gbps InfiniBand™ network. The computational capacity of *abe.lustre* is roughly equivalent to that of *c1.xlarge*, and the comparative performance to estimate the virtualization overhead on AmEC2.

**Allocation of Storage**

Our experiments required storage for application executables, input data, and intermediate and output data. Application executables are pre-installed on the execution site, while input data are copied from local disk or an archive and staged on the execution site, and output data are copied from the execution site to a local disk.

For AmEC2, executables were installed in the Virtual Machine (VM) images, intermediate and output data were written to a local partition, and input data were stored on *Elastic Block Store* (EBS) volumes. EBS is a Storage Area Network-like, replicated, block-based storage service that supports volumes between 1 GB and 1 TB in size that appear as standard, unformatted block devices when attached to an AmEC2 resource. As such, EBS volumes can be formatted as standard UNIX file systems and used like an ordinary disk attached to one instance; no sharing between instances is possible. EBS was chosen to store input data for three reasons:

- Storing input data in the cloud obviates the need to transfer input data repeatedly (and is economical because data transfer is more expensive than data storage).
- We can access the data as if they were on a local disk without packaging it in the VM image and without compromising performance.
- EBS simplifies the experimental setup by allowing us to reuse the same volume for multiple experiments by detaching it from one volume and re-attaching to another.

For Abe, all application executables and input files were stored in the Lustre file system. For *abe.local* experiments, the input data were copied to a local partition before running the workflow, and all intermediate and output data were written to the same local partition. For *abe.lustre*, all intermediate and output data were written to the Lustre file system.

**Software Organization**

The execution environment took advantage of existing workflow management and scheduling tools. All workflows on AmEC2 and Abe were planned and executed using the *Pegasus Workflow Management System* [8] with *DAGMa*n (http://www.cs.wisc.edu/condor/dagman/) and *Condo*r [9]. Pegasus is used to transform abstract workflow descriptions into concrete processing plans, which are then executed through DAGMan to manage task dependencies, and through Condor to manage task execution. Our execution environments were set up to provide, as far as was possible, equivalent environments on AmEC2 and Abe. A submit host running at ISI was used to send workflow tasks to AmEC2 and Abe worker nodes.

*Amazon EC2*: Two virtual machine images were used for AmEC2 worker nodes: one for 32-bit resource types and one for 64-bit resource types. Both images were based on the standard Fedora Core 8 images provided by AmEC2, to which we added Condor, Pegasus and other miscellaneous packages required to compile and run the selected applications. The 32-bit image was 2.1 GB (773 MB compressed) and the 64-bit image was 2.2 GB (729 MB compressed). The images did not include any application-specific configurations, so we were able to use the same set of images for all experiments. The images were stored in Amazon S3, an object-based, replicated storage service that supports simple PUT and GET operations to store and retrieve files.

*Abe:* Globus [10] and Corral (http://pegasus.isi.edu/glidein/latest/) were used to deploy Condor *glideins* [11] that started Condor daemons on the Abe worker nodes, which in turn contacted the submit host and were used to execute workflow tasks. This approach creates an execution environment on Abe that is equivalent to the AmEC2 environment.

Table 2: Summary of the Processing Resources on Amazon EC2 and the Abe High Performance

| Type | Architecture | CPU | Cores | Memory | Network | Storage | Price |
|---|---|---|---|---|---|---|---|
| Amazon EC2 | | | | | | | |
| m1.small | 32-bit | 2.0-2.6 GHz Opteron | 1-2 | 1.7 GB | 1-Gbps Ethernet | Local | $0.10/hr |
| m1.large | 64-bit | 2.0-2.6 GHz Opteron | 2 | 7.5 GB | 1-Gbps Ethernet | Local | $0.40/hr |
| m1.xlarge | 64-bit | 2.0-2.6 GHz Opteron | 4 | 15 GB | 1-Gbps Ethernet | Local | $0.80/hr |
| c1.medium | 32-bit | 2.33-2.66 GHz Xeon | 2 | 1.7 GB | 1-Gbps Ethernet | Local | $0.20/hr |
| c1.xlarge | 64-bit | 2.0-2.66 GHz Xeon | 8 | 7.5 GB | 1-Gbps Ethernet | Local | $0.80/hr |

| Abe Cluster | | | | | | | |
|---|---|---|---|---|---|---|---|
| abe.local | 64-bit | 2.33 GHz Xeon | 8 | 8 GB | 10-Gbps InfiniBand | Local | … |
| abe.lustre | 64-bit | 2.33 GHz Xeon | 8 | 8 GB | 10-Gbps InfiniBand | Lustre™ | … |

## 3. EVALUATING APPLICATIONS ON THE AMAZON EC2 CLOUD: RESULTS

### 3.1 Performance Comparison Between Amazon EC2 and the Abe High Performance Cluster

Figure 2 compares the runtimes of the Montage, Broadband and Epigenome workflows on all the Amazon EC2 and Abe platforms listed in Table 2. Runtime in this context refers to the total amount of wall clock time in seconds from the moment the first workflow task is submitted until the last task completes. These runtimes exclude the following:

- The time required to install and boot the VM, which typically averages between 70 and 90 seconds (AmEC2 only)
- The time glidein jobs spend waiting in the queue, which is highly dependent on the current system load (Abe only)
- The time to transfer input and output data, which varies with the load on the Wide Area Network (WAN). In our experiments, we typically observed bandwidth on the order of 500-1000KB/s between EC2 and the submit host in Marina del Rey, CA.

This definition of runtime (also known as "makespan") enables comparison of the performances of Amazon EC2 and Abe on a one-to-one basis. The similarities between the specifications of *c1.xlarge* and *abe.local* provide an estimate of the virtualization overhead for each application on AmEC2.

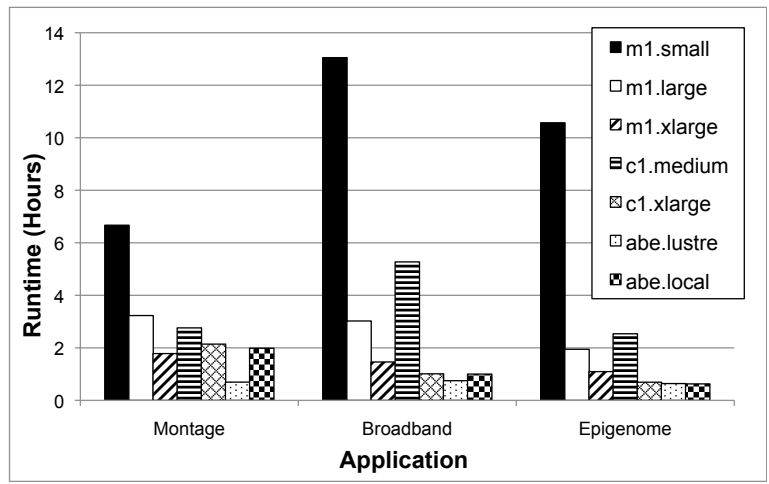

Figure 2: The processing times for the Montage, Broadband and Epigenome workflows on the Amazon EC2 cloud and Abe (see text). The legend identifies the processors.

**Montage (I/O bound)**
The best performance was achieved on *m1.xlarge* resource type. This resource type has double the memory of the other types, and the extra memory is used by the Linux kernel for the file system buffer cache to reduce the amount of time the application spends waiting for I/O. This is particularly beneficial for an I/O intensive application like Montage. Reasonably good performance was achieved on all except *m1.small*, which is much less powerful than the other AmEC2 resource types. The EC2 *c1.xlarge* type is nearly equivalent to *abe.local* and delivered nearly equivalent performance (within 8%), which indicates that the virtualization overhead does not seriously degrade performance.

The most important result of Figure 2 is a demonstration of the performance advantage of high-performance parallel file systems for an I/O-bound application. While the AmEC2 instances are not prohibitively slow, the processing times on *abe.lustre* are nevertheless nearly three times faster than the fastest AmEC2 machines. A parallel file system could in principle be constructed from the raw components available in AmEC2, but the cost would be prohibitive and would be of little practical value without a high-speed network, for the I/O bottleneck would shift from the disk to the network. AmEC2 would need to deploy both a parallel file system and a high-speed interconnect to make dramatic performance upgrades, and the cost of such a major architectural upgrade would presumably be passed on to customers.

**Broadband (Memory bound)**
For a memory bound application such as Broadband, the processing advantage of the parallel file system largely disappears: *abe.lustre* offers only slightly better performance than *abe.local*. And *abe.local*'s performance is only 1% better than *c1.xlarge*, so virtualization overhead is essentially negligible. For a memory-intensive application like Broadband, AmEC2 can achieve nearly the same performance as Abe as long as there is more than 1 GB of memory per core. If there is less, then some cores must sit idle to prevent the system from running out of memory or swapping. Broadband performs the worst on *m1.small* and *c1.medium*, the machines with the smallest memories (1.7 GB). This is because *m1.small* has only half a core, and *c1.medium* can only use one of its two cores because of memory limitations.

**Epigenome (CPU bound)**
As with Broadband, the parallel file system in Abe provides no processing advantage: processing times on *abe.lustre* were only 2% faster than on *abe.local*. Epigenomes' performance suggests that virtualization overhead may be more significant for a CPU-bound application: the processing time for *c1.xlarge* was some 10% larger than for *abe.local*. As might be expected, the best performance for Epigenome was obtained with those machines having the most cores.

### 3.2 Cost-analysis of Running Workflow Applications on Amazon EC2

AmEC2 itemizes charges for hourly use of all of its resources, including charges for
- Resources, including the use of VM instances and processing
- Data storage, including the cost of virtual images in S3 and input data in S3.
- Data transfer, including charges for transferring input data into the cloud, and
- Transferring output data and log files between the submit host and AmEC2.

**Resource Cost**
AmEC2 charges different hourly rates for its resources, with generally higher rates as the processor speed, number of cores and size of memory increase. Column 8 of Table 2 summarizes the hourly rates for the machines used in this experiment, current as of Summer 2009. They vary widely, from $0.10/hr for *m1.small* to $0.80/hr for *m1.xlarge* and *c1.xlarge* (Note: after these experiments were run Amazon reduced its prices, so the values reported here are higher than they would be if the experiments were repeated today). Figure 3 shows the resource cost for the workflows whose performances were given in Figure 2.

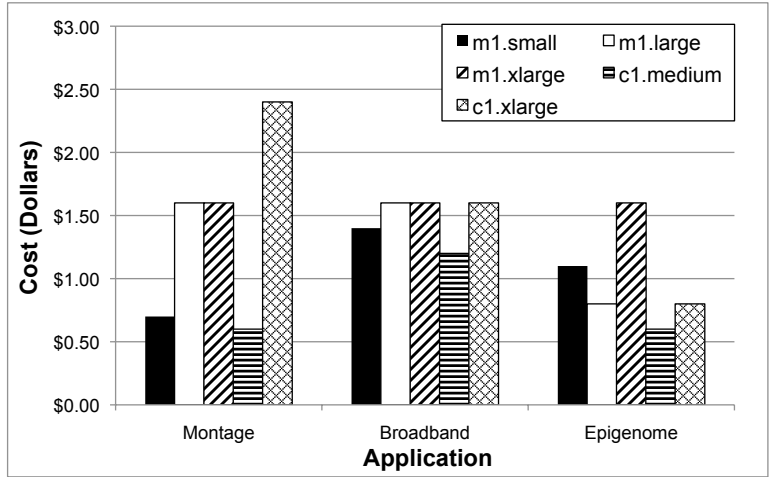

Figure 3: The processing costs for the Montage, Broadband and Epigenome workflows for the Amazon EC2 processors given in the legend

The Figure clearly shows the trade-off between performance and cost for **Montage**. The most powerful processor, *c1.xlarge*, offers a 3-threefold performance advantage over the least powerful, m1.small, but at 5 times the cost. The most cost-effective solution is *c1.medium*, which offers performance of only 20% less than *m1.xlarge* but at 5-times lower cost.

For **Broadband**, the picture is quite different. Processing costs do not vary widely with machine, so there is no reason to choose other than the most powerful machines. Similar results apply to **Epigenome:** the machine offering the best performance, *c1.xlarge*, is the second cheapest machine.

**Storage Cost**

Storage cost consists is made up of cost to store VM images in S3, and the cost of storing input data in EBS. Both S3 and EBS use fixed monthly charges for the storage of data, and charges for accessing the data; these vary according to the application. The rates for fixed charges are $0.15 per GB-month for S3, and $0.10 per GB-month for EBS. The variable charges are $0.01 per 1,000 PUT operations and $0.01 per 10,000 GET operations for S3, and $0.10 per million I/O operations for EBS.

The 32-bit image used for the experiments in this paper was 773 MB, compressed, and the 64-bit image was 729 MB, compressed, for a total fixed cost of $0.22 per month. In addition, there was 4616 GET operations and 2560 PUT operations for a total variable cost of approximately $0.03. The fixed monthly cost of storing input data for the three applications is shown in Table 3. In addition, there were 3.18 million I/O operations for a total variable cost of $0.30.

Table 3: Monthly Storage Cost

| Application | Input Volume | Monthly Cost |
|---|---|---|
| Montage | 4.3 GB | $0.66 |
| Broadband | 4.1 GB | $0.66 |
| Epigenome | 1.8 GB | $0.26 |

**Transfer Cost**

In addition to resource and storage charges, AmEC2 charges $0.10 per GB for transfer into the cloud, and $0.17 per GB for transfer out of the cloud. Tables 4 and 5 show the transfer sizes and costs for the three workflows. "Input" is the amount of input data to the workflow, "output" is the amount of output data, and "logs" refers to the amount of logging data that is recorded for workflow tasks and transferred back to the submit host. The cost of the protocol used by Condor to communicate between the submit host and the workers is not included, but it is estimated to be much less than $0.01 per workflow.

Table 5 summarizes the input and output sizes and costs. While data transfer costs for Epigenome and Broadband are small, for Montage they are larger than the processing and storage costs using the most cost-effective resource type. Given that scientists will almost certainly need to transfer products out of the cloud, transfer costs may prove prohibitively expensive for high-volume products. While the cost of transferring input data can be amortized by storing them in the cloud, the cost of transferring output data may be more difficult to reduce.

Table 4: Data transfer sizes per workflow on Amazon EC2

| Application | Input | Output | Logs |
|---|---|---|---|
| Montage | 4,291 MB | 7,970 MB | 40 MB |
| Broadband | 4,109 MB | 159 MB | 5.5 MB |
| Epigenome | 1,843 MB | 299 MB | 3.3 MB |

Table 5: The costs of transferring data into and out of the Amazon EC2 cloud

| Application | Input | Output | Logs | Total |
|---|---|---|---|---|
| Montage | $0.42 | $1.32 | $<0.01 | $1.75 |
| Broadband | $0.40 | $0.03 | $<0.01 | $0.43 |
| Epigenome | $0.18 | $0.05 | $<0.01 | $0.23 |

AmEC2 end-users of I/O intensive applications producing large outputs need to give serious consideration to the tradeoff between storage cost and transfer-out cost. Users may transfer input data for each run of a workflow, or transfer input data once, and storing them in the cloud. The choice of approach to employ depend on how many times the data will be used, how long the data will be stored, and how frequently the data will be accessed. In general, storage is more cost-effective for input data that are reused often and accessed frequently, and transfer is more cost-effective if data will be used only once. For the workflows tested in this paper, the monthly cost to store input data is only slightly more than the cost to transfer it once. Therefore, for these applications, it may more cost-effective to store the input data rather than transfer the data for each workflow, but a cost-benefit analysis should be performed in each case as part of selecting a processing and storage strategy. This is especially the case for users wishing to store large data sets in the cloud, for the costs may become prohibitive. For example, the cost of storing the 2MASS image survey (10-TB in three bands) would be $61K over three years, exclusive of transfer costs, and yet could be stored locally on a disk farm costing only $15K.

### 3.3 Conclusions

- Virtualization overhead on AmEC2 is generally small, but most evident for CPU-bound applications.
- The resources offered by AmEC2 are generally less powerful than those available in high-performance clusters and generally do not offer the same performance. This is particularly the case for I/O–bound applications, whose performance benefits greatly from the availability of parallel file systems. This advantage essentially disappears for CPU and memory bound applications.
- End-users should understand the resource usage of their applications and undertake a cost benefit study of the resources offered them to establish a processing and storage strategy. They need to take into account factors such as:
    - AmEC2 itemizes charges for resource usage, data transfer and storage, and the impact of these costs should be evaluated
    - For I/O-bound applications, the most expensive resources are not necessarily the most cost-effective
    - Data transfer costs can exceed the processing costs for I/O-bound applications.
    - AmEC2 offers no cost benefits over locally hosted storage, and is generally more expensive, but does

## 4. IMAGE MOSAICS AND PROVENANCE MANAGEMENT

### 4.1 Production of Mosaics

In the production steps shown in Figure 1, the files output by one step become the input to the subsequent step. That is, the reprojected images are used as input to the background rectification. This rectification itself consists of several steps that fit a model to the differences between flux levels of each image, and in turn the rectified, reprocessed images are input to the co-addition engine. Thus the production of an image mosaic actually generates a volume of data that is substantially greater than the volume of the mosaic. Table 6 illustrates this result for two use cases that return 3-color mosaics from the Two Micron All Sky Survey (2MASS) images (see http://www.ipac.caltech.edu/2mass/). One is a 6 deg sq mosaic of ρ Oph and the second is an All Sky mosaic. The table makes clear that the volume of intermediate products exceeds the mosaic size by factors of 30 to 50. The Infrared Processing and Analysis Center (IPAC) hosts an on-request image mosaic service (see Section 4.3) that delivers mosaics of user-specified regions of the sky, and it currently receives 25,000 queries per year. Were mosaics of the size of the ρ Oph mosaic processed with such frequency, the service would produce 3.8 PB of data each year. Such volumes are clearly too high to archive.

Table 6: Estimates of Files Generated in the Production of Image Mosaics. See text for an explanation of "Annual Volume."

|  | ρ Oph 6 deg sq | All Sky Mosaic |
|---|---|---|
| # input images | 4,332 | 4,121,439 |
| # comp. steps | 25,258 | 24,030,310 |
| # intermediate products | 67,300 | 61,924,260 |
| Size of intermediate products | 153 GB | 126 TB |
| Mosaic Size | 2.4 GB | 4 TB |
| Annual Volume | 3.8 PB | … |

### 4.2 The Scientific Need To Reprocess Mosaics

Montage makes three assumptions and approximations that affect the quality of the mosaics:

- Re-projection involves redistributing the flux from the input pixel pattern to the output pixel pattern. Montage uses a fast, custom algorithm that approximates tangent plane projections (that is, geometric projections of the celestial sphere onto a tangent plane from a center of projection at the center of the sphere) as polynomial approximations to the pixel pattern on the sky, which can produce small distortions in the pixel pattern of the mosaic.

- There is no physical model of the sky background that predicts its flux as a function of time and wavelength. Montage assumes that the sky background is only significant at the lowest spatial frequencies, and rectifies the flux at these frequencies to a common level across all the input images. This approximation can confuse background flux with an astrophysical source present at the same frequencies, such as extended diffuse emission in a nebula or dust cloud.

- Co-additions of the re-projected, rectified images are weighted to not take into account outliers due to e.g. residual cosmic ray hits.

Users have two options in investigating the impact of these three factors, and both involve knowing the provenance of the mosaics:

1. Analyze the output from intermediate steps to understand how the features in the mosaic originate.

2. Replace modules with implementations of new algorithms, such as a custom background rectification, and reprocess the mosaic.

### 4.3 Information Needed In Provenance Records

Column 1 of Table 7 lists all the information needed to specify a provenance record for an image mosaic. To illustrate the current quality of provenance recording, column 2 describes the provenance information that is made available to users by an on-line, on-request image mosaic service at http://hachi.ipac.caltech.edu:8080/montage/. This service is hosted at IPAC, and returns mosaics of 2MASS, Sloan Digital Sky Survey (SDSS) and Digitized Sky Surveys at Space Telescope (DSS) images. When processing is complete, users are directed to a web page that contains links to the mosaic and to processing information. Column 2 of table 7 refers to the contents of these pages.

The only information that is permanently recorded are the runtime parameters that specify the properties of the image mosaic - the coordinate system, projection, spatial sampling and so on - written as keywords in the mosaic file header. The file itself, as well as log files and the traceability to the input images, are deleted after 72 hours (but these can be reproduced if the user has a record of the specifications of the mosaic requested). There is no record of the execution environment, and the algorithm and software information are described in the project web page, and presume that users know where to find them and that the web pages do not become stale.

Table 7: Comparison of Required and Recorded Provenance Information

| Information | Recorded In On-Request Service |
| --- | --- |
| **Algorithms** | |
| Algorithm Design Documents | Accessible from Montage web page |
| Algorithm Version | Accessible from Montage web page |
| **Execution Environment** | |
| Specific hardware | No |
| OS and Version | No |
| Process Control and Management Tools | No |
| **Software** | |
| Software Source Code, version | Accessible from Montage web page |
| Software Build Environment, version | Accessible from Montage web page |
| Compiler, version | Accessible from Montage web page |
| Dependencies and versions | Accessible from Montage web page |
| Test Plan Results | Accessible from Montage web page |
| **Runtime** | |
| Parameters | Included in output files |
| Input files, version | Retained for 72 hours after completion of job |
| Output Files, Log Files | Retained for 72 hours after completion of job |

## 4.4 Experiments in Recording Provenance Information

The previous section reveals an obviously unsatisfactory state of affairs. We have therefore investigated how astronomers may take advantage of methodologies already under development in other fields to create and manage a permanent store of provenance records for the Montage engine. When complete, these investigations are intended to deliver an operational provenance system that will enable replication of any mosaic produced by Montage.

## 4.5 Characteristics of Applications and Provenance Management

The design of Montage is well suited for the creation of provenance records, as follows (see [12] for more details):

- It is deterministic; that is, processing a common set of input files will yield the same output mosaic.
- It is component based, rather than monolithic.
- It is self-contained and requires, e.g., no distributed services.
- It runs on all common hardware platforms.
- It inputs data in self-describing standard formats.
- Its input data are curated and served over the long term.
- 

## 4.6 Capturing the Provenance of Montage Processing

Many custom provenance systems are embedded in processing environments, which offer the benefits of efficient collection of self-contained provenance records, but at the cost of ease of interoperation with other provenance systems. Given that Montage can be run as a pipeline, it too can employ such a system [12]. In this paper, we will report instead on efforts to leverage an existing methodology to create a standardized provenance store that can interoperate with other applications. The methodology is the Provenance Aware Service Oriented Architecture (PASOA) [13], an open source architecture already used in fields such as aerospace engineering, organ transplant management, and bioinformatics. In brief, when applications are executed they produce documentation of the process that is recorded in a *provenance store*, essentially a repository of provenance documents and records. The store is housed in a database so that provenance information can be queried and accessed by other applications .

In our investigation, Montage was run with the Pegasus framework [8]. Pegasus was developed to map complex scientific workflows onto distributed resources. It operates by taking the description of the processing flow in Montage (the abstract workflow) and mapping it onto the physical resources that will run it, and records this information in its logs. It allows Montage to run on multiple environments and takes full advantage of the parallelization inherent in the design. Pegasus has been augmented with PASOA to create a provenance record for Montage in eXtended Markup Language (XML) that captures the information identified in Table 7. We show a section of this XML structure below, captured during the creation of a mosaic of M17:

```xml
<?xml version="1.0" encoding="ISO-8859-1"?>
<invocation xmlns="http://vds.isi.edu/invocation" xmlns:xsi="http://www.w3.org/2001/XMLSchema-instance"
xsi:schemaLocation="http://vds.isi.edu/invocation http://vds.isi.edu/schemas/iv-1.10.xsd" version="1.10"
start="2007-03-26T16:25:54.837-07:00" duration="12.851" transformation="mShrink:3.0" derivation="mShrink1:1.0"
resource="isi_skynet" hostaddr="128.9.233.25" hostname="skynet-15.isi.edu" pid="31747" uid="1007" user="vahi"
     gid="1094"
group="cgt" umask="0022">
<prejob start="2007-03-26T16:25:54.849-07:00" duration="5.198" pid="31748">
  <usage utime="0.010" stime="0.030" minflt="948" majflt="0" nswap="0" nsignals="0" nvcsw="688" nivcsw="4"/>
  <status raw="0"><regular exitcode="0"/></status>
```

```xml
<statcall error="0">
  <!-- deferred flag: 0 -->
  <file name="/nfs/home/vahi/SOFTWARE/space_usage">23212F62696E2F73680A686561646572</file>
  <statinfo mode="0100755" size="118" inode="20303958" nlink="1" blksize="32768" blocks="8"
      mtime="2007-03-26T11:06:24-07:00"
atime="2007-03-26T16:25:52-07:00" ctime="2007-03-26T11:08:12-07:00" uid="1007" user="vahi" gid="1094" group="cgt"/>
  </statcall>
  <argument-vector>
  <arg nr="1">PREJOB</arg>
  </argument-vector>
</prejob>
<mainjob start="2007-03-26T16:26:00.046-07:00" duration="2.452" pid="31752">
  <usage utime="1.320" stime="0.430" minflt="496" majflt="8" nswap="0" nsignals="0" nvcsw="439" nivcsw="12"/>
  <status raw="0"><regular exitcode="0"/></status>
  <statcall error="0">
  <!-- deferred flag: 0 -->
  <file name="/nfs/home/mei/montage/default/bin/mShrink">7F454C4601010100000000000000000</file>
  <statinfo mode="0100755" size="1520031" inode="1900596" nlink="1" blksize="32768" blocks="2984"
mtime="2006-03-22T12:03:36-08:00" atime="2007-03-26T14:16:36-07:00" ctime="2007-01-11T15:13:06-08:00"
 uid="1008" user="mei" gid="1008" group="mei"/>
  </statcall>
  <argument-vector>
  <arg nr="1">M17_1_j_M17_1_j.fits</arg>
  <arg nr="2">shrunken_M17_1_j_M17_1_j.fits</arg>
  <arg nr="3">5</arg>
  </argument-vector>
</mainjob>
```

---

Our experiments with PASOA have been successful and a next step will be to deploy it as part of an operational system.

### 4.7 Applications in Earth Sciences and Oceanography

While the work described above is an advanced experimental stage, Earth Sciences and Oceanography projects have for a number of years exploited operational provenance management systems [7]. We would suggest that astronomy has much to learn from these projects. Here we describe two examples, one involving an integrated pipeline, and one involving a complex data system that uses many instruments collecting a complex and dynamic data set.

***Example 1: The Moderate Resolution Imaging Spectroradiometer (MODIS):*** An instrument launched in 1999 aboard the Terra platform, MODIS scans the Earth in 36 bands every two days. The raw ("level 0") data are transformed into calibrated, geolocated products ("level 1B"), which are then aggregated into global data products ("level 2") that are the primary map. The raw data are archived permanently, but the level 1B data are much too large to archive. These data are retained for 30-60 days only. Consequently, the MODIS archive records all the process documentation needed to reproduce the Level 1B data from the raw satellite data. The process documentation includes the algorithms used, their versions, the original source code, a complete description of the processing environment and even the algorithm design documents themselves [14].

***Example Two: The Monterey Bay Aquarium Shore Side Data System (SSDS):*** For the past four years, the SSDS has been used to track the provenance of complex data sets form many sources [15]. Oceanographers undertake campaigns that involve taking data from multiple sources - buoys, aircraft, underwater sensors, radio-sondes and so on. These instruments measure quantities such as salinity and amount of chlorophyll. These data are combined with published data including satellite imagery in simulations to predict oceanographic features, such as seasonal variations in water levels. The SSDS was developed to track the provenance of the data measured in the campaigns in standardized central repository. Scientists use SSDS to track back from derived data products to the metadata of the sensors including their physical location, instrument and platform. The system automatically populates metadata fields, such as the positions of instruments on moving platforms.

### 4.8 Conclusions

Tracking the provenance of data products will assume ever-growing importance as more and larger data sets are made available to astronomers.

- Methodologies such as PASOA are in use in aerospace and bioinformatics applications and show great promise for providing provenance stores for astronomy.

- Earth Science projects routinely track provenance information. There is much that astronomy can learn from them.

- There is also an effort in the provenance community to standardize on a provenance model [16], intended to foster interoperability between provenance systems and spur on the development of generic provenance capture and query tools.

## 5. FINAL REMARKS

Our study has shown that cloud computing offers a powerful and cost-effective new resource for scientists, especially for compute and memory intensive applications. For I/O bound applications, however, high-performance clusters equipped with parallel file systems and high performance networks do offer superior performance. End-users should perform a cost-benefit study of cloud resources as part of their usage strategy. Experiments with existing provenance management, already is use on other fields, show great promise in astronomy as well.